\documentclass[aps,pre,onecolumn,showkeys,nofootinbib]{revtex4-2}

\usepackage{amsmath,amssymb}
\usepackage{graphicx}
\usepackage[caption=false]{subfig} 
\usepackage{hyperref}
\usepackage{float}
\usepackage{enumitem}
\usepackage{physics}
\usepackage{fancyhdr}
\usepackage{etoolbox}
\usepackage{microtype}
\usepackage{siunitx} 
\usepackage[dvipsnames]{xcolor}

\usepackage[margin=1in]{geometry}

\hypersetup{
  colorlinks=true,
  linkcolor=MidnightBlue,
  citecolor=MidnightBlue,
  urlcolor=MidnightBlue,
  pdfauthor={Ferris Moser},
  pdftitle={Investigating the Phase Space Dynamics of Hamiltonian Systems through the Origin--Fate Map}
}

\newcommand{\papertitle}{Investigating the Phase Space Dynamics of Hamiltonian Systems by the Origin-Fate Map}
\newcommand{\paperauthor}{Ferris Moser}
\newcommand{\paperaffil}{Nonlinear Dynamics and Chaos Group\\ Department of Mathematics and Applied Mathematics\\ University of Cape Town}
\newcommand{\paperdate}{\today}

\makeatletter
\patchcmd{\@dottedtocline}{\hfil}{\hspace{0.8em}\hfil}{}{}
\makeatother

\begin{document}

\thispagestyle{empty}
\begin{titlepage}
  \vspace*{-1.5cm}

  \begin{center}
    {\LARGE\bfseries \papertitle \par}
    \vspace{0.8em}

    {\large \paperauthor \par}
    \vspace{0.3em}
    {\normalsize \paperaffil \par}
    \vspace{0.1em}
    {\normalsize \paperdate \par}
    \vspace{1.0em}
  \end{center}

  \begin{center}
  \begin{minipage}{0.92\linewidth}
    \small
    \noindent\textbf{Abstract}\par
    \vspace{0.2em}
    \noindent\rule{\linewidth}{0.4pt}\par
    \vspace{0.3em}
    \setlength{\parskip}{0.4em}
    We investigate phase space transport in a two-dimensional stretched caldera potential using the Origin–Fate Map (OFM) framework, complemented by Lagrangian Descriptor (LD) analysis.  
    The caldera potential, a model for reaction dynamics with multiple exit channels, is adjusted by a stretching factor $\lambda$ controlling the directional bias of the four‐saddle landscape.  
    Several OFMs are constructed for two Poincaré surfaces of section using forwards and backwards symplectic integration to assign each initial condition a channel of origin (entrance) and fate (exit).  
    Our results reproduce the highly symmetric $\lambda=1.0$ patterns reported in \textit{Hillebrand \textit{et al.}}, \textit{Phys.\ Rev.\ E}~108,~024211~(2023) \cite{Hillebrand2023} and reveal, for smaller $\lambda$, a pronounced channel imbalance, figure–eight transport loops, and complex mixed–channel chaotic regions.  
    Long‐time integrations show a reduction of trapped regions with boundaries exhibiting self–similarity under deep zoom with fractal-like structures.  
    A high–resolution zoom in the $\lambda=0.4$ OFM uncovers intricate lobe structures and string-like boundaries.  
    The LD fields computed on this domain, and their gradient magnitude, show the invariant manifolds that govern the transport. The thresholded ridges align precisely with the OFM boundaries.
    \vspace{0.3em}
    \noindent\rule{\linewidth}{0.4pt}
  \end{minipage}
  \end{center}

  \vspace{0.5em}
  \begin{center}
    \small \textit{Keywords:} Origin--Fate Map, Lagrangian Descriptors, Caldera Potential, Phase Space Transport, Hamiltonian Chaos
  \end{center}

  \vspace{0.8em}

  \begin{center}
      \includegraphics[width=0.65\textwidth]{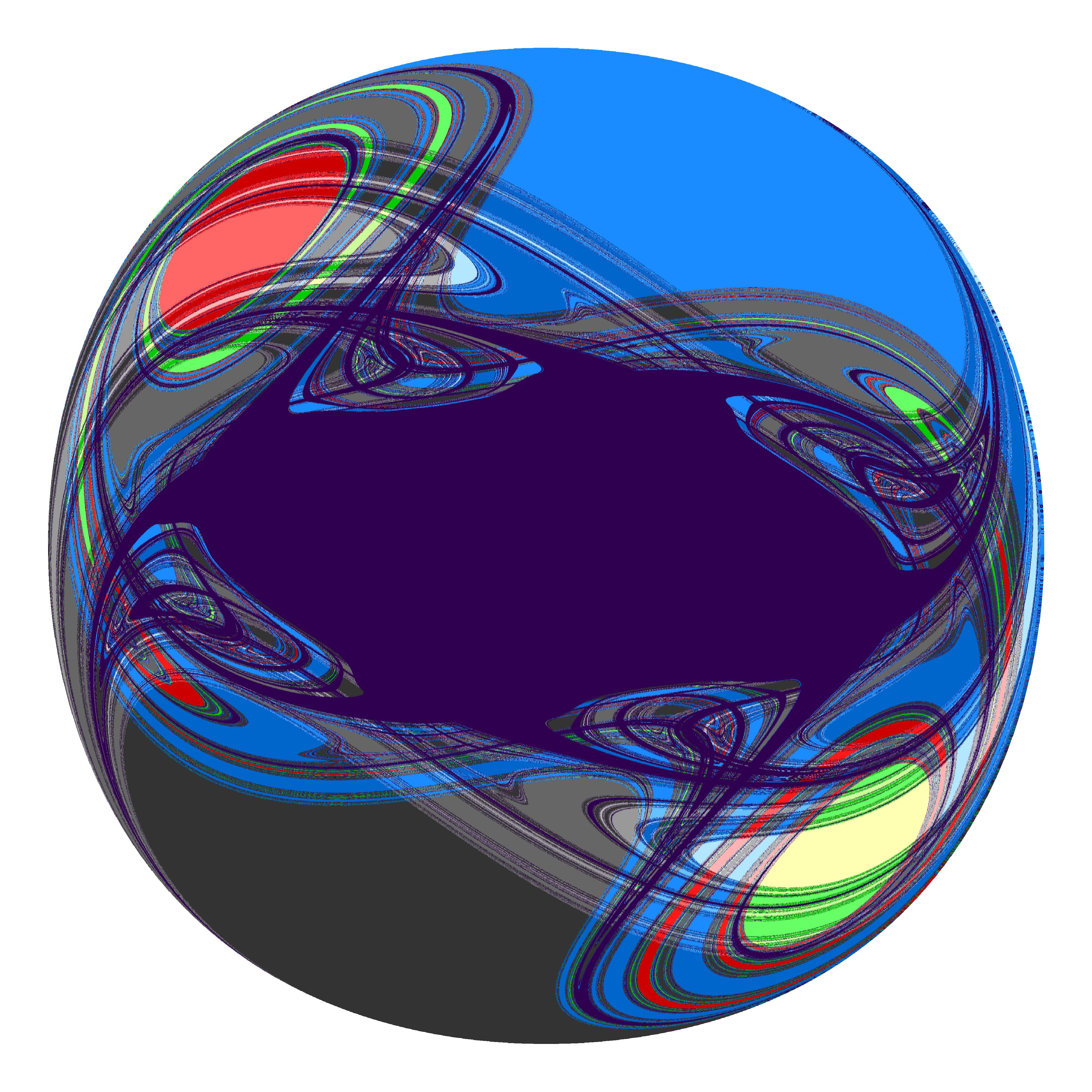}
  \end{center}

  \vfill
\end{titlepage}

\setcounter{page}{1}
\pagestyle{fancy}

{\Large\bfseries Table of Contents}
\vspace{0.5em}

\tableofcontents

\newpage
\onecolumngrid
\fancypagestyle{plain}{
  \fancyhf{}
  \fancyhead[R]{Ferris Moser}
  \fancyhead[L]{Investigating Phase Space via the OFM}
  \fancyfoot[C]{\thepage}
}
\pagestyle{plain}

\section{Introduction}
Hamiltonian systems are a foundational framework for classical mechanics, describing the evolution of dynamical systems through a set of differential equations derived from the scalar function, which is known as the Hamiltonian. This set of systems has constant energy when the Hamiltonian function does not explicitly depend on time, and shows a number of behaviours which range from integrable to chaotic dynamics, depending on the potential energy landscape \cite{Wiggins1992}. In fields ranging from molecular dynamics to celestial mechanics, it is imperative that the global organisation of the phase space structures be understood. The phase space dynamics are governed by the Hamiltonian equations, which show the evolution of momentum and position variables over time. 

A main focus in these studies is understanding how the initial conditions evolve over time and in which exit and entrance channels they are from and lead to within a region of phase space. Some of the classic approaches to this include using Poincaré surfaces of section (PSS) and the computation of stable and unstable manifolds of periodic orbits. The more recent tools include Lagrangian Descriptors (LDs) to reveal phase space transport structures based on finite-time orbit differences \cite{Mancho2013,Lopesino2017}. The LDs are very useful when it comes to high-dimensional systems, as they provide a scalar field that indicates barriers to mixing and transition. 

The tricky part of the methods is extracting the transport information that is useful to us. The \emph{Origin–Fate Map} (OFM) that was introduced in \cite{Hillebrand2023}, gives a direct classification of transport through Hamiltonian systems with multiple exit channels. By integrating each orbit forwards and backwards in time, the initial condition can be assigned a label showing where it originates (the origin) and exits (the fate). This allows us to construct the representation of phase space transport and picture manifolds, lobes and symmetry breaking without deliberately computing them. 

In their study, Hillebrand et al.\ used a two-degree-of-freedom system. They applied their OFM framework to it using a modified caldera potential, which is typically used to model chemical reactions. The system contains four exit channels, with the parameter $\lambda$ controlling the stretching of the potential along the $x$-axis. As $\lambda$ gets smaller, the OFM region starts to be flooded by chaotic transport and broken dynamical matching. This makes the caldera ideal for testing OFM-based transport classification.

This study aims to reproduce and build upon the results created and presented in \cite{Hillebrand2023}. In particular, the OFM computation is implemented using the 4th-order symplectic integrator ABA864 \cite{McLachlan1995} as it can achieve a higher accuracy over longer time steps than lower-order methods.
The original OFM structures are recreated and then built upon by using a dense set of $\lambda$ values, allowing us to generate a sequence of OFMs that show the change in the orbits as the potential stretches. This method is also used to examine the initial conditions in both the Poincaré surface of section with $y = 1.88409$ (from the original study) and in $y = 0.3$ with $p_y < 0$, which lead to distinctly different OFMs and perfectly overlapping forwards and backwards integrated orbit paths.

In order to build upon the OFM analysis we also include Poincaré sections to show the intersection of orbits within a fixed phase space plane as well as use LDs to find the transport features and set manifolds of the OFM. These methods offer us a platform to view and track how particles move through a landscape and how symmetry breaking influences their fate.

This report is organised as follows. Section~\ref{sec:theory} introduces the theoretical background of the Hamiltonian dynamics and caldera potential. Section~\ref{sec:methods} explains the numerical setup, the integration method, and the general techniques to simulate the OFMs, LDs, and Poincaré sections. Section~\ref{sec:results} shows the structure of the OFM for different initial conditions, which include zoom-enhanced sections of the OFM, as well as comparing the orbit evolutions. Lastly, the conclusions and future directions are discussed in Section~\ref{sec:conclusions}.

\section{Theoretical Background}
\label{sec:theory}
\subsection{Hamiltonian Systems}

Hamiltonian systems are a class of dynamical systems that encapsulate the theoretical progression in time of physical processes in a variety of fields, which include classical mechanics and chemical dynamics. The scalar function: 
\begin{equation}
\mathcal{H}(q, p) = T(p) + V(q),
\label{eq:Hamiltonian}
\end{equation}
called the Hamiltonian, represents the system's total energy as the sum of its kinetic ($T(p)$) and potential energy ($V(q)$) and generates the corresponding Hamiltonian dynamics through Hamilton’s equations.

In \eqref{eq:Hamiltonian} $q = (q_1, \ldots, q_n)$ are generalised coordinates, $p = (p_1, \ldots, p_n)$ are their conjugate momenta.

Hamilton’s equations:
\begin{equation}
\dot{q}_i = \frac{\partial \mathcal{H}}{\partial p_i}, \qquad \dot{p}_i = -\frac{\partial \mathcal{H}}{\partial q_i}, \qquad i = 1,2,\ldots,n,
\label{Hamiltons}
\end{equation}
determine the time evolution of the system. Eq~\eqref{Hamiltons} represent a system of $2n$ first-order differential equations. These equations ensure the conservation of total energy and the symplectic structure of phase space.

A fundamental property of Hamiltonian systems is Liouville's theorem, which states that under time evolution, the volume of phase space is conserved. With this, combined with the energy conservation and the ability to reverse time, phase space becomes intricate yet orderly. When phase space corresponds to systems with more than two degrees of freedom, the phase space becomes a diverse mixture of chaotic layers, stochastic seas and islands of stability \cite{Wiggins1992,Ott2002}. 

An essential model in the study of non-linear Hamiltonian dynamics is the Hénon–Heiles system \cite{Henon1964}, introduced initially to model the motion of a star around a galactic centre. Its Hamiltonian is given by: 

\begin{equation}
\mathcal{H}(x,y,p_x,p_y) = \frac{1}{2}(p_x^2 + p_y^2) + \frac{1}{2}(x^2 + y^2) + x^2 y - \frac{1}{3}y^3,
\label{eq:henon_heiles}
\end{equation}

\noindent where $(x,y)$ are Cartesian coordinates and $(p_x,p_y)$ their conjugate momenta.  
This system transitions from regular to chaotic motion as the total energy $\mathcal{H}$ increases.  
Motion is confined within a bounded potential well for energies below the escape threshold $\mathcal{H} = \frac{1}{8}$.  
Above this threshold, three symmetric exit channels open, allowing orbits to escape to infinity.  
The appearance of both islands of stability and chaotic sea regions makes it a perfect example of mixed phase space dynamics, and its invariant manifold structure is well-studied in the context of open Hamiltonian systems \cite{Wiggins1992,Ott2002}.  
Throughout this work, the Hénon–Heiles model serves as a conceptual reference.

In this project, two-degree-of-freedom Hamiltonian systems of the form:
\begin{equation}
\mathcal{H}(x, y, p_x, p_y) = \frac{1}{2}(p_x^2 + p_y^2) + V(x, y).
\label{eq:hamiltonian}
\end{equation}
are used with Cartesian coordinates $(x, y)$ and their corresponding momenta $(p_x, p_y)$. The associated equations of motion are given by:
\begin{equation}
\begin{cases}
\dot{x} = p_x, & \dot{y} = p_y, \\
\dot{p}_x = -\dfrac{\partial V}{\partial x}, & \dot{p}_y = -\dfrac{\partial V}{\partial y}.
\end{cases}
\end{equation}

Changing the initial conditions even slightly in these Hamiltonian systems can result in entirely different orbits. This is especially brought about when the potential landscape includes wells or exit channels. Analyses of the structure that guides the transportation of these orbits, such as the invariant manifolds, are critical for understanding how a given particle will move between regions of phase space and through which channels they will enter or exit \cite{MacKay1994, Dellago1998}.

Multiple techniques have been developed to reveal these structures in phase space, which include Poincaré sections, LDs and the OFM. These methods offer different perspectives on the evolution of orbits, which all form a theoretical basis for analysing complex Hamiltonian transport, which is explored more in the following sections. 

\subsection{Caldera Potential}

The caldera potential is a model used in theoretical chemistry to represent reaction dynamics in systems which include multiple exit channels and transition states. The name originates from the Spanish word for cauldron, which is reflected by the structure of the surface map with its well-shaped central region and exit channels surrounding it. Originally used to study the competition between direct and indirect reaction pathways, the caldera potential is now widely used for investigating non-linear transport, symmetry breaking and chaotic scattering in lower-dimensional Hamiltonian systems \cite{Collins2014, Ezra2009}.

In this study, a modified version of the two-dimensional caldera potential is considered, as used by Hillebrand et al.~\cite{Hillebrand2023}, to investigate the structure of phase space and the classification of transport behaviour via the OFM. The potential is defined as:
\begin{equation}
V(x, y) = c_1(\lambda^2 x^2 + y^2) + c_2 y - c_3(\lambda^4 x^4 + y^4 - 6 \lambda^2 x^2 y^2),
\label{eq:caldera_potential}
\end{equation}
where $\lambda$ controls the potential stretching along the $x$-axis. The coefficient $c_1$ sets the harmonic confinement strength, $c_2$ is a linear y-bias term which introduces a vertical asymmetry, and $c_3$ controls the influence of non-linear terms that generate saddles, lobes, and chaotic transport structures.

\begin{figure}[H]
    \centering
    \includegraphics[width=0.8\textwidth]{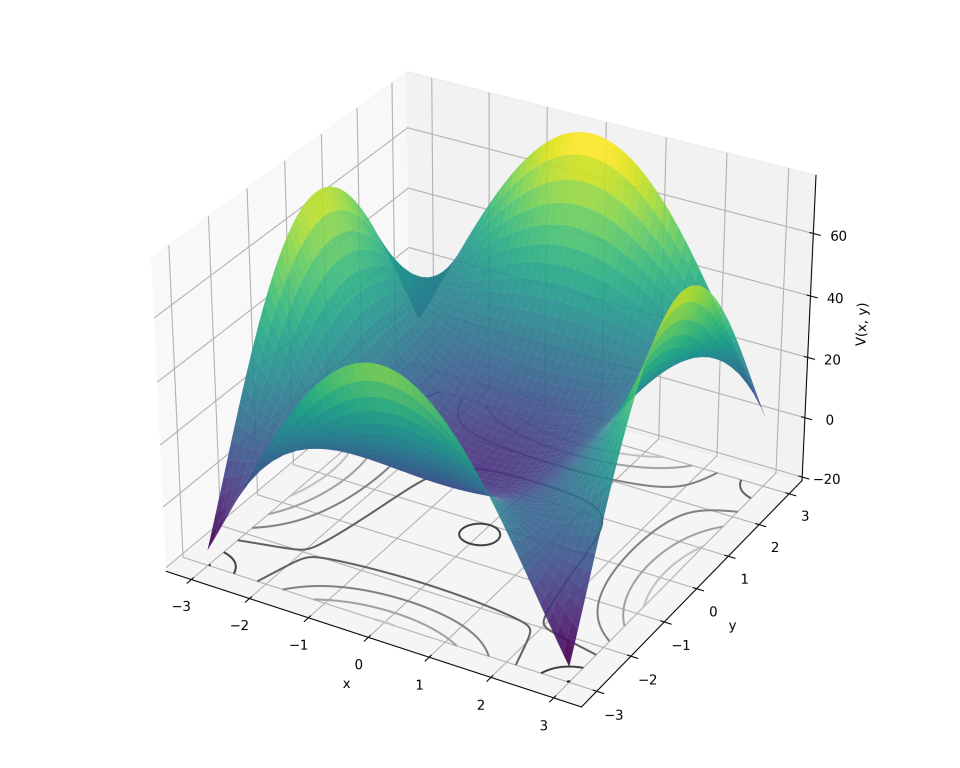}
    \caption{The unstretched ($\lambda = 1$) caldera potential energy surface $V(x,y)$ as defined in Eq.~(\ref{eq:caldera_potential}), showing the central well and four saddle‐point exit channels located in each quadrant.}
    \label{fig:caldera3d}
\end{figure}

This modified potential $V(x,y)$ creates four exit channels that are located in each quadrant of the $(x, y)$ plane, shown in Fig.~\ref{fig:caldera3d}. The well in the middle connects to the exits through saddle regions and these regions greatly change shape when the value of $\lambda$ is varied. For larger $\lambda$ values the system is condensed and oval shaped. As $\lambda$ gets smaller, the system stretches out along the $x$-axis and larger regions of chaos appear.

The ability to modify the caldera potential is especially useful when exploring transport phenomena in phase space. By fixing the total energy above the saddle thresholds, particles can escape through the exit channels. Small variations in initial conditions then determine which channel an orbit follows, highlighting the system’s sensitivity to initial conditions. This sensitivity makes the caldera an ideal setting for studying invariant manifolds, transport behaviours, and lobe dynamics. The OFM framework reveals these features in the sections that follow.

\section{Numerical Methods}
\label{sec:methods}
\subsection{Symplectic Integrators}
\label{sec:methods:SI}

Hamiltonian systems evolve in phase space according to Hamilton's equations, preserving a symplectic structure~\cite{Wiggins1992,Ott2002}. This property is responsible for conserving energy and phase space volume over time. When simulating these systems, especially over long periods of time, usual integrators (like Runge–Kutta) can lead to the conserved quantities, like the total energy, changing when they shouldn't~\cite{McLachlan1995}. Symplectic integrators avoid this change by ensuring that the system's geometric structure is preserved, thus making them ideal for simulations of phase space dynamics that have higher fidelity.  \\

Standard integrators, such as Runge–Kutta, often accumulate energy errors over time, leading to drift and distortion of the true dynamics. In contrast, symplectic integrators like ABA864 keep the energy error tightly bounded. The accuracy of an integrator in conserving the Hamiltonian can be quantified through the \textit{relative energy error}, defined as
\[
\Delta H_{\mathrm{rel}}(t) = \frac{|H(t) - H(0)|}{|H(0)|},
\]
which measures the deviation of the instantaneous energy from its initial value as a fraction of the total energy. As shown in Fig.~\ref{fig:energyerror}, the relative energy error remains below $10^{-8}$ and oscillates around $10^{-9.5}$, even over long integration times. This makes the ABA864 integration particularly well-suited for long-time simulations where preserving the system’s qualitative behaviour, such as energy conservation, is essential.

In this work, we use the ABA864 symplectic integrator developed by McLachlan~\cite{McLachlan1995}, due to its exceptionally low energy error at fixed time steps compared to standard integrators. It is a symmetric 8-stage composition method that preserves the system’s Hamiltonian structure and maintains a bounded energy deviation over long-time simulations in the form of:

\begin{figure}[H]
    \centering
    \includegraphics[width=0.7\textwidth]{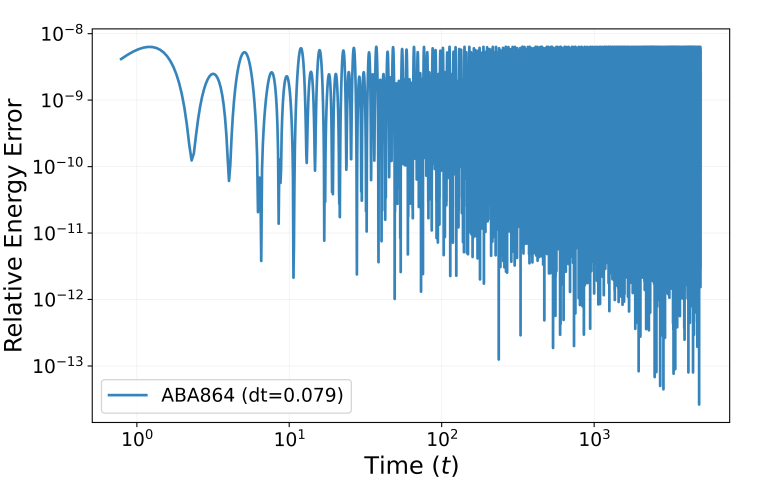}
    \caption{Relative energy error over time using the ABA864 integrator for the time evolution of the orbit governed by the Hénon–Heiles Hamiltonian, Eq.~\eqref{eq:hamiltonian}, with initial conditions $\left(x_0,\, y_0,\, p_{y,0}\right) = \left(0,\, 0.1,\, 0.1\right)$, and $p_{x,0}$ determined from the energy constraint $\mathcal{H}(x_0, y_0, p_{x,0}, p_{y,0}) = \tfrac{1}{8}$. The integration time step is $\Delta t = 0.079$.}

    \label{fig:energyerror}
\end{figure}

\[
S(\Delta t)
= A_{a_1}B_{b_1}
  A_{a_2}B_{b_2}
  A_{a_3}B_{b_3}
  A_{a_4}B_{b_4}
  A_{a_4}B_{b_3}
  A_{a_3}B_{b_2}
  A_{a_2}B_{b_1}
  A_{a_1},
\]
where each $A_{a_i}$ step advances the position variables (the kinetic step), and each $B_{b_i}$ step updates the momentum variables (the potential step). 
This symmetric composition ensures time–reversibility and high–order accuracy while preserving the symplectic geometry of the Hamiltonian system. \\[1em]

The ABA864 method achieves sixth–order accuracy using the following full–precision coefficients:
\[
\begin{aligned}
a_1 &=  0.0711334264982231,  &\quad b_1 &=  0.1830836874721972, \\
a_2 &=  0.2411534279566401,  &\quad b_2 &=  0.3107828598985749, \\
a_3 &=  0.5214117617728148,  &\quad b_3 &= -0.0265646185119588, \\
a_4 &= -0.3336986162276780,  &\quad b_4 &=  0.0653961422823734. \\
\end{aligned}
\]
The coefficients are applied in a palindromic sequence:
\[
(a_1, a_2, a_3, a_4, a_4, a_3, a_2, a_1),
\qquad
(b_1, b_2, b_3, b_4, b_3, b_2, b_1),
\]
ensuring symmetry and exact time reversibility. \\[0.5em]

The update rule that evolves the system forward by one time step $\Delta t$ is:

\begin{align}
&\text{(A-step, kinetic update)} &&
x \leftarrow x + a_i\,\Delta t\, p_x, &
y \leftarrow y + a_i\,\Delta t\, p_y, \nonumber\\[0.5em]
&\text{(B-step, potential update)} &&
p_x \leftarrow p_x - b_i\,\Delta t\, \frac{\partial V}{\partial x}(x, y), &
p_y \leftarrow p_y - b_i\,\Delta t\, \frac{\partial V}{\partial y}(x, y).
\end{align}

To ensure that the integrator is sufficiently accurate for our simulations, we use a fixed integration time step of $\Delta t = 0.079$ and simulate an orbit in the Hénon–Heiles system with energy $E = \tfrac{1}{8}$. An initial condition $(x_0, y_0, p_y) = (0.0,\, 0.1,\, 0.1)$ is selected, and the corresponding $p_x$ value is computed to satisfy the energy constraint exactly. The energy error is then monitored over $t = 5000$ units of time. 

In particular, in Fig.~\ref{fig:energyerror}, we consider the orbit initialized at $(x_0, y_0, p_x, p_y) = (0.0,\, 0.1,\, 0.0977,\, 0.1)$, corresponding to the target energy $E = 1/8$. This orbit illustrates the near-perfect energy conservation achieved by the ABA864 symplectic integrator, where the relative energy error remains bounded throughout the simulation without any systematic drift. This level of stability is crucial for analysing long-term phase space transport and for constructing accurate PSSs and OFMs.

\subsection{Poincaré Sections}
\label{sec:methods:poincare}

The PSS is a classic tool in Hamiltonian dynamics for reducing the dimensionality of systems phase space and visualising their underlying structure. By taking the intersection of the phase space with a fixed hyperplane and recording the crossing of the orbits, the PSS transforms the continuous orbits into a map that reveals structures like the chaotic seas and islands of stability \cite{Wiggins1992,Ott2002}.

This study considers the surface defined by $x = 0$ and keeps only intersections for which $p_x > 0$. These crossings correspond to particles moving to the right through the $x = 0$ plane in a particular way. For a given energy $H$, the momenta must satisfy the constraint:
\begin{equation}
\mathcal{H} = \frac{1}{2}(p_x^2 + p_y^2) + V(x, y),
\end{equation}
where $V(x, y)$ is the Hénon–Heiles potential defined in Eq.~\eqref{eq:henon_heiles}. In particular, we consider initial conditions, $(y, p_y)$ values which are sampled uniformly, and the corresponding $p_x$ is computed by solving the energy constraint. Only real and positive solutions for $p_x$ are kept, ensuring crossings in a particular way .

Each orbit is integrated using the ABA864 symplectic integrator, which is described in Section~\ref{sec:methods}, ensuring that the computed intersections preserve the phase space structure over longer time periods. After each time step, we check for a sign change in $x(t)$, and linear interpolation is used to estimate the exact intersection point with $x = 0$.

\begin{figure}[H]
    \centering
    \includegraphics[width=0.45\textwidth]{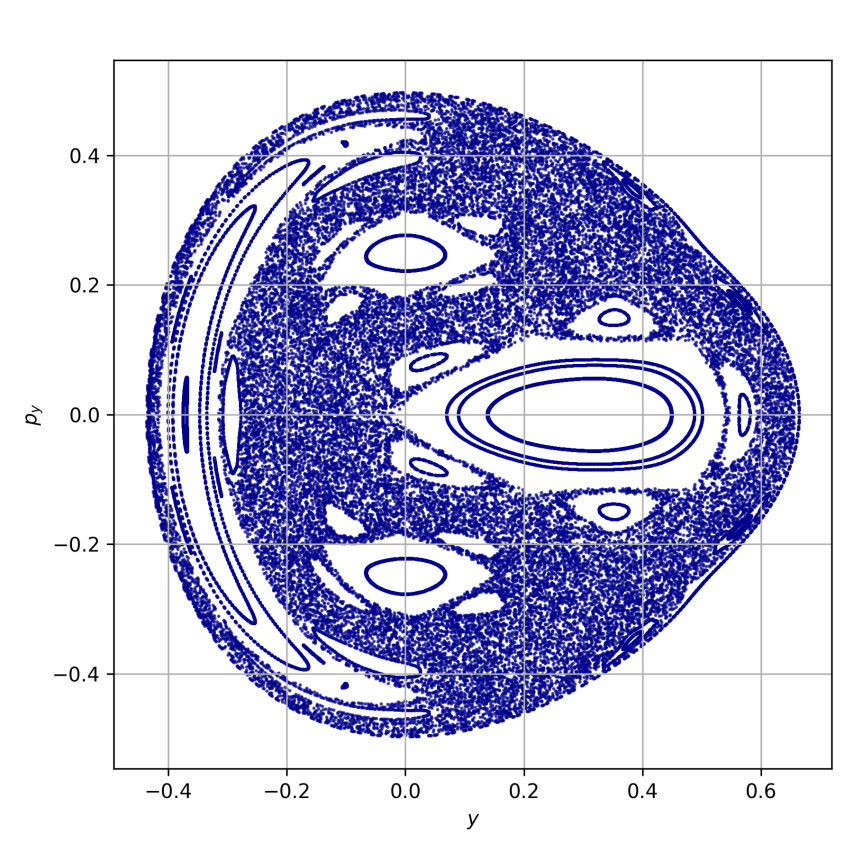}
    \caption{Poincaré section of the system \eqref{eq:henon_heiles} for $x = 0$ and $p_x > 0$ computed using the ABA864 integrator at fixed energy $H = 1/8$.}
    \label{fig:poincare}
\end{figure}

To visualise the structure for the Poincaré map, the points $(y, p_y)$ are plotted for all real and positive intersections. Fig.~\ref{fig:poincare} reveals the intricate phase space geometry, where the coexistence of invariant tori and stochastic transport regions becomes apparent through the mix of regular and chaotic structures. The shape and extent of these regions are controlled by the total energy $H$.

\subsection{Lagrangian Descriptors}
\label{sec:methods:ld}

LDs are a scalar field method used to show invariant structures in dynamical systems by calculating the arc length of orbits over a given time window. It was used in fluid dynamics initially to track fluid flow patterns but was later adapted into Hamiltonian systems \cite{Mancho2013, Lopesino2017}. The LDs can reveal the so-called skeleton of phase space, which includes the stable and unstable manifolds, chaotic seas, and islands of stability, without requiring prior knowledge of the system’s geometric properties. 

Given an initial condition $\mathbf{x}_0$ at time $t = 0$, the LD is computed as
\begin{equation}
LD(\mathbf{x}_0) = \int_{-\tau}^{\tau} \left\| \frac{d\mathbf{x}(t)}{dt} \right\| dt \approx \sum_{i=-N}^{N-1} \| \mathbf{x}_{i+1} - \mathbf{x}_i \|,
\end{equation}
where $\mathbf{x}_i$ are phase-space points along the orbit obtained by numerical integration and $\tau$ is the half–integration window. In our implementation, we use a symmetric window with $\tau=20$ and a fixed time step $dt=0.05$, so that $N=\tau/dt$ steps are taken forwards and backwards in time. Low LD values typically indicate regular motion on tori, whereas sharp spatial changes or higher LD derivative values reflect the presence of manifolds and separatrices. This is supported by the analysis presented in \cite{Lopesino2017} which connects LD features to invariant manifolds.

We evaluate LDs on the PSS defined by $x=0$ with $p_x>0$ at fixed energy $H=1/8$, as is standard for Hamiltonian systems \cite{Wiggins1992, Ott2002}. A uniform grid of initial conditions $(y,p_y)\in[-0.5,0.7]\times[-0.6,0.6]$ with resolution $n=800$ is constructed. For each $(y,p_y)$, the compatible $p_x$ is obtained from the energy constraint:
\[
\frac{p_x^2}{2}+\frac{p_y^2}{2} + \underbrace{\Big[\tfrac{1}{2}(x^2+y^2)+x^2y-\tfrac{1}{3}y^3\Big]}_{V(x,y)}\Bigg|_{x=0}
= \mathcal{H},
\]
i.e. $p_x^2 = 2\big[\mathcal{H} - \tfrac{1}{2}p_y^2 - \tfrac{1}{2}y^2 + \tfrac{1}{3}y^3\big]$, and points with $(p_x)^2<0$ are discarded. Each orbit is then integrated forwards and backwards with the ABA864~\cite{McLachlan1995} symplectic integrator, and the sum of the arc–length in $(y,p_y)$ yields the LD field on the grid.

\begin{figure*}[t]
    \centering
    \includegraphics[width=\textwidth]{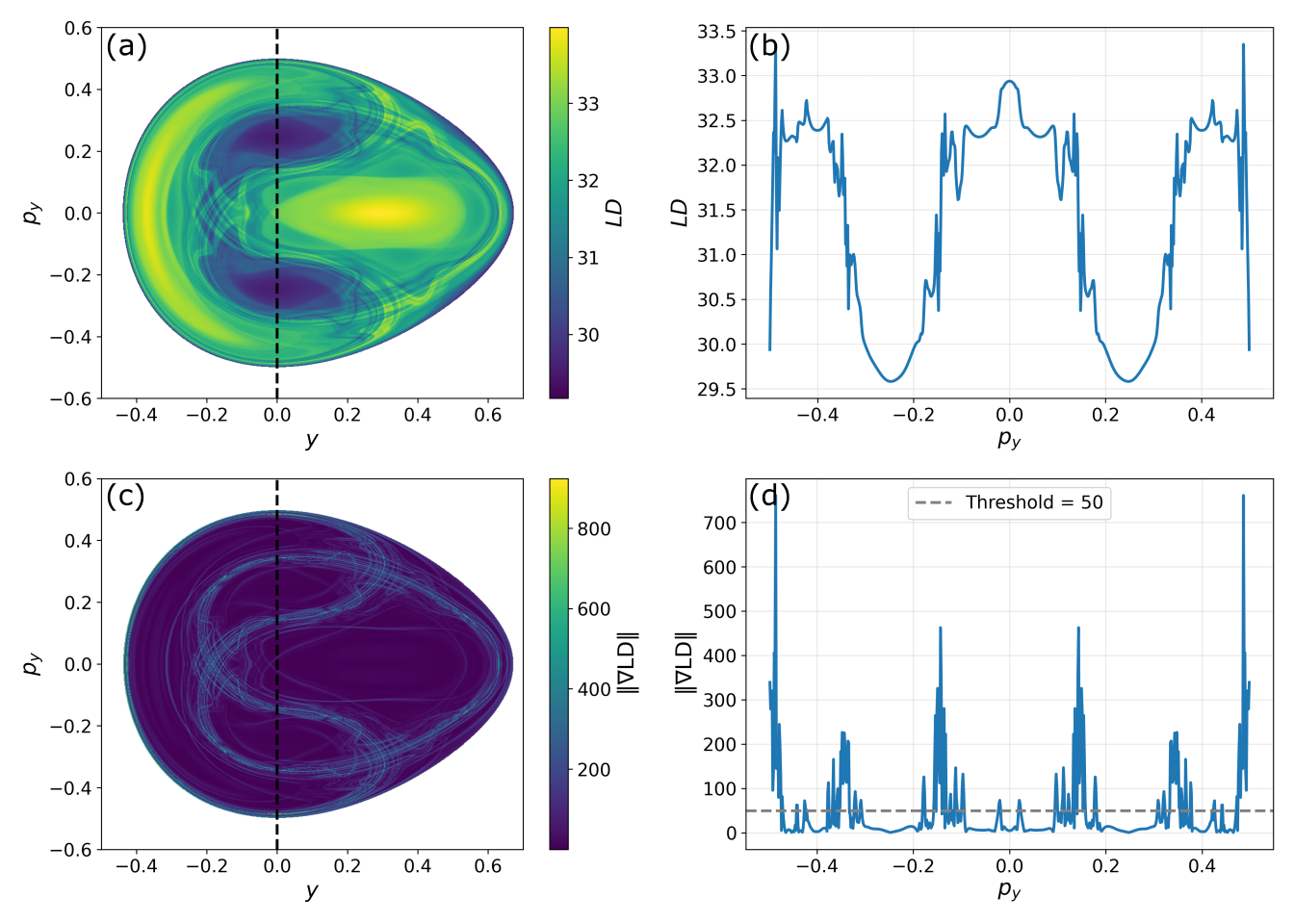}
    \caption{Lagrangian Descriptor computation on the Poincaré section ($x=0$, $p_x>0$, $H=1/8$) for the HH system \eqref{eq:henon_heiles}. Integration uses the ABA864 scheme with integration time step $dt=0.05$, $\tau=20$, grid $800\times800$ over $(y,p_y)\in[-0.6,0.6]$. (a) LD field; (b) 1D LD profile along $y=0$ indicated by a dashed vertical line in (a); (c) gradient magnitude $|\nabla \mathrm{LD}|$; (d) $|\nabla \mathrm{LD}|$ profile along $y=0$ indicated by a vertical dashed line in (c). The dashed line indicates the threshold $|\nabla \mathrm{LD}|=50$ used to identify manifold structures.}
    \label{fig:ld_poincare_top}
\end{figure*}

The resulting LD field (Fig.~\ref{fig:ld_poincare_top}a) shows smooth regions corresponding to regular motion and thinner thread-like zones where manifolds appear. To enhance these ridges, we compute the gradient magnitude of the LD field,
\[
\|\nabla \mathrm{LD}(y,p_y)\| \;=\; \sqrt{\Big(\tfrac{\partial\,\mathrm{LD}}{\partial y}\Big)^2 + \Big(\tfrac{\partial\,\mathrm{LD}}{\partial p_y}\Big)^2},
\]
and, on the discrete grid, we approximate it with second–order central differences (grid spacings $\Delta y$ and $\Delta p_y$):
\[
\Big(\tfrac{\partial\,\mathrm{LD}}{\partial y}\Big)_{i,j} \approx \frac{\mathrm{LD}_{i,\,j+1}-\mathrm{LD}_{i,\,j-1}}{2\,\Delta y}, 
\qquad
\Big(\tfrac{\partial\,\mathrm{LD}}{\partial p_y}\Big)_{i,j} \approx \frac{\mathrm{LD}_{i+1,\,j}-\mathrm{LD}_{i-1,\,j}}{2\,\Delta p_y},
\]
\[
\|\nabla \mathrm{LD}\|_{i,j} \approx \sqrt{
\left[\frac{\mathrm{LD}_{i,\,j+1}-\mathrm{LD}_{i,\,j-1}}{2\,\Delta y}\right]^2+
\left[\frac{\mathrm{LD}_{i+1,\,j}-\mathrm{LD}_{i-1,\,j}}{2\,\Delta p_y}\right]^2 }.
\]
In addition to these 2-D fields (Fig.~\ref{fig:ld_poincare_top}c), we extract a one-dimensional profile of LD along the slice $y=0$ (Fig.~\ref{fig:ld_poincare_top}b), revealing local minima and maxima linked to the crossing of invariant tori and manifolds. The corresponding $\|\nabla \mathrm{LD}\|$ profile along the same slice (Fig.~\ref{fig:ld_poincare_top}d) is compared to a fixed threshold of $50$ to identify points belonging to the manifold set.

\begin{figure}[H]
    \centering
    \includegraphics[width=0.5\textwidth]{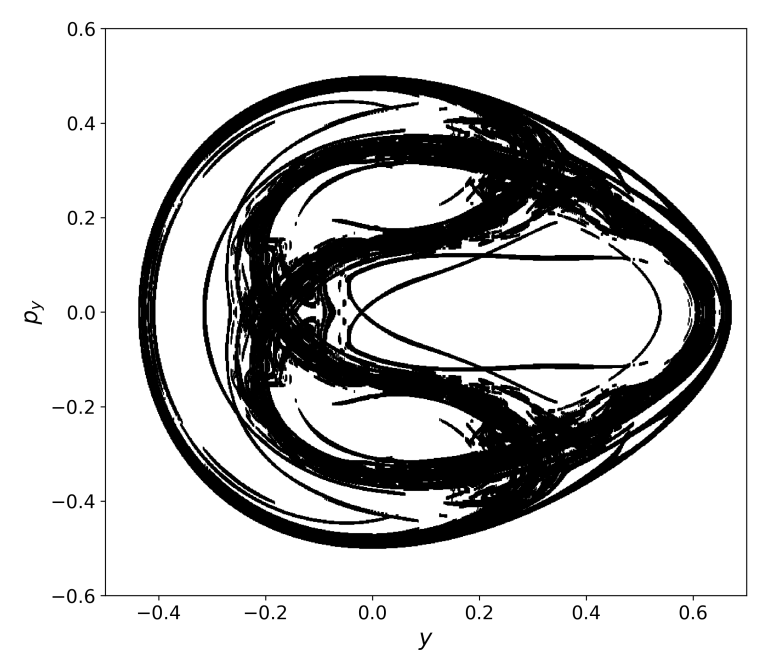}
    \caption{Manifold set extracted from regions where $|\nabla \mathrm{LD}| > 50$, in Fig.~\ref{fig:ld_poincare_top}. Thin ridges denote invariant manifolds and separatrices that bind transport regions.}
    \label{fig:ld_poincare_mask}
\end{figure}

Applying this fixed threshold to the full $\|\nabla \mathrm{LD}\|$ map generates the binary manifold mask shown in Fig.~\ref{fig:ld_poincare_mask}, where the thin ridges align with stable/unstable manifolds and separatrices that bound transport regions. This view makes the transport partitions and guidelines for orbits that will be compared directly with OFM structures in Sec.~\ref{sec:methods:ofm}.

\subsection{The Origin--Fate Map}
\label{sec:methods:ofm}

The OFM is a tool for visualising transport behaviour in Hamiltonian systems with multiple exit channels. It assigns a label to a given orbit based on which channel it enters (origin) and from which it exits (fate) a particular region of phase space (e.g., a potential well). The method was introduced by Hillebrand et al.~\cite{Hillebrand2023} as an extension of the traditional fate-map concept to time-reversible Hamiltonian systems, enabling a complete classification of orbits in terms of both entry and exit channels. It builds on earlier developments in chaotic transport and invariant-manifold theory~\cite{Wiggins1992,RomKedar1990,Bleher1988}.

In our implementation, initial conditions are sampled on a 2D grid in the $(x, p_x)$ plane at fixed $y = y_0$, and momenta $p_y$ are computed to satisfy the total energy constraint:
\begin{equation}
H(x, y, p_x, p_y) = E
\end{equation}

Only real-valued solutions for $p_y$ are considered to ensure consistency of initial conditions. Each orbit then is integrated both forwards and backwards in time using the ABA864 integrator \cite{McLachlan1995} for a fixed integration time $\tau$ or until it escapes the central well of the caldera potential. Escape is defined as the orbit exceeding a threshold of $|y| > 6$ as in \cite{Hillebrand2023}, and each escaping orbit is assigned a symbolic channel index based on the quadrant through which it exits.

\begin{itemize}[noitemsep]
    \item Channel 1: Upper Left ($x < 0$, $y > 6$)
    \item Channel 2: Upper Right ($x > 0$, $y > 6$)
    \item Channel 3: Lower Right ($x > 0$, $y < -6$)
    \item Channel 4: Lower Left ($x < 0$, $y < -6$)
\end{itemize}

This applies to both the forwards and backwards integration. Each initial condition is assigned an \emph{origin} and a \emph{fate} channel. A unique symbolic index is computed as:

\begin{equation}
\text{Index} = 4 \cdot (\text{origin} - 1) + (\text{fate} - 1) + 1,
\end{equation}

Each initial condition is assigned both an \emph{origin} and a \emph{fate} channel. Since there are four exit channels, this produces $4 \times 4 = 16$ possible origin–fate combinations. In addition, orbits that fail to escape in at least one time direction (forwards or backwards) are classified as trapped and placed in a separate class. Altogether, this results in 17 distinct categories, which are represented in the OFM by a 17-colour map encoding the full transport behaviour of the system.

\begin{figure}[H]
    \centering
    \includegraphics[width=0.45\textwidth]{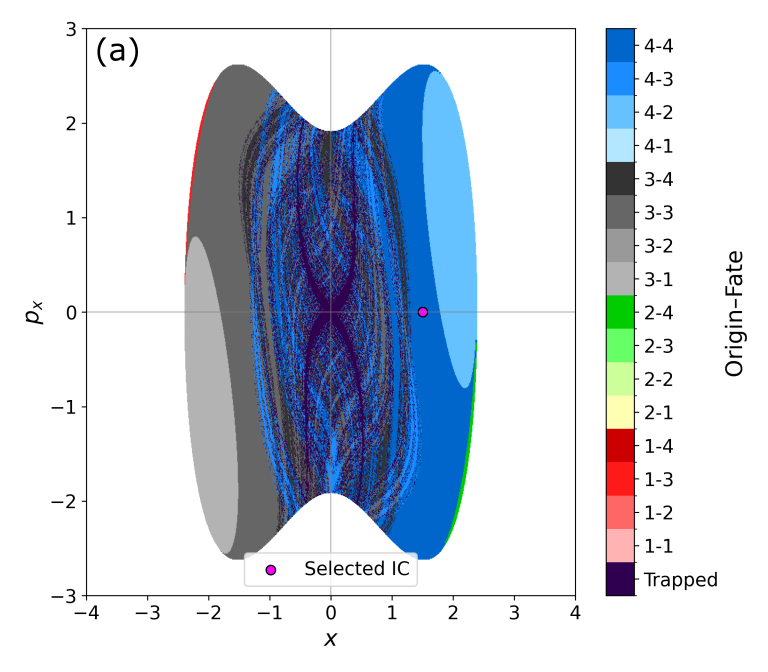}
    \includegraphics[width=0.45\textwidth]{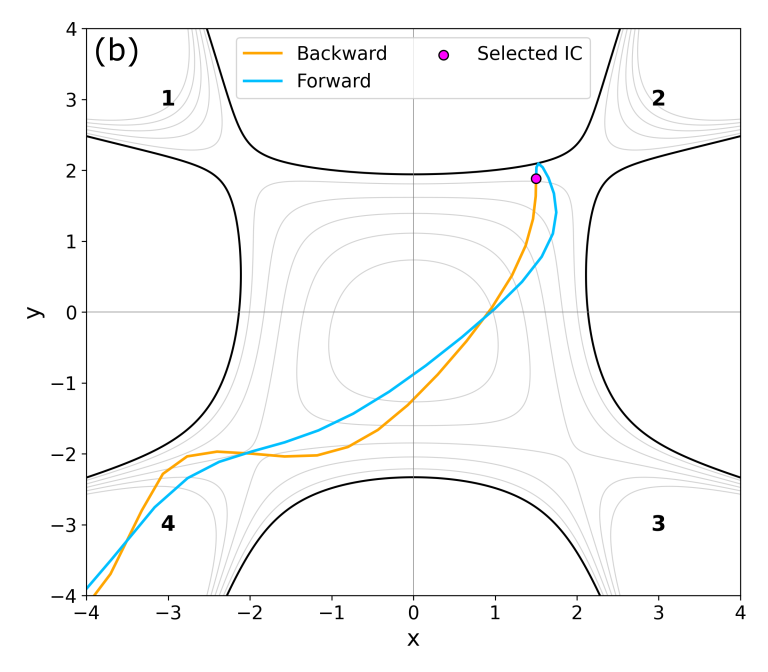}
    \caption{(a) The OFM computed for $\lambda = 1.0$, $E = 29$, and $\tau = 20$, on the section $y = 1.88409$ with $p_y > 0$, showing the outcome of forwards and backwards integration for each initial condition. (b) Configuration space view of the forwards (blue) and backwards (orange) orbits of the magenta point in panel (a). The black contours indicate the zero-velocity curves $V(x,y) = E$, which bound the energetically accessible region.}
    \label{fig:ofm_tau20}
\end{figure}

Figure~\ref{fig:ofm_tau20}a shows a representative OFM at $\lambda = 1.0$ and energy $E = 29$ on the Poincaré slice $y = 1.88409$ with $p_y > 0$, following the setup of Hillebrand et al.~\cite{Hillebrand2023}. Each colour corresponds to a unique origin–fate combination. Panel (b) shows the forwards and backwards evolution of an orbit created by a selected initial condition marked on the OFM.

The OFM enables the symbolic interpretation of transport processes and gives insight into underlying dynamical structures, such as the invariant manifolds and lobe dynamics \cite{Mancho2013, Lopesino2017}. Particularly when it comes to how the different fate regions correspond to stable manifolds, while the origin boundaries reflect manifolds. The way that these structures evolve with $\lambda$ and $\tau$ is explored further in Section~\ref{sec:results}.


\section{Results}
\label{sec:results}
\subsection{Origin--Fate Maps for the PSS defined by $y = 1.88409$, $p_y > 0$}
\label{sec:results:case1}
The OFM provides a global representation of how orbits launched from a fixed initial condition distribute among the available exit channels of the caldera potential.  
Here we consider the PSS $y = 1.88409$ with $p_y > 0$, so that orbits enter the central well from below.  
The procedure for OFM computation follows the method described in Sec.~\ref{sec:methods:ofm}, where forwards and backwards integrations are used to assign an ``origin'' and a ``fate'' channel index to each initial condition.

\begin{figure*}[t]
    \centering
    \includegraphics[width=0.48\textwidth]{Pics/OFM_lambda_1.000.png}
    \includegraphics[width=0.48\textwidth]{Pics/Trajectory_lambda_1.000.png} \\
    \includegraphics[width=0.48\textwidth]{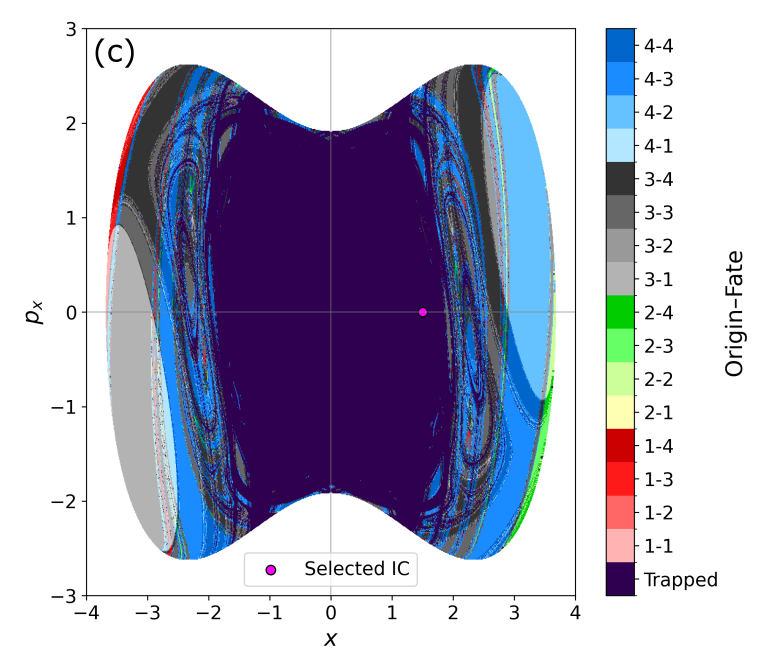}
    \includegraphics[width=0.48\textwidth]{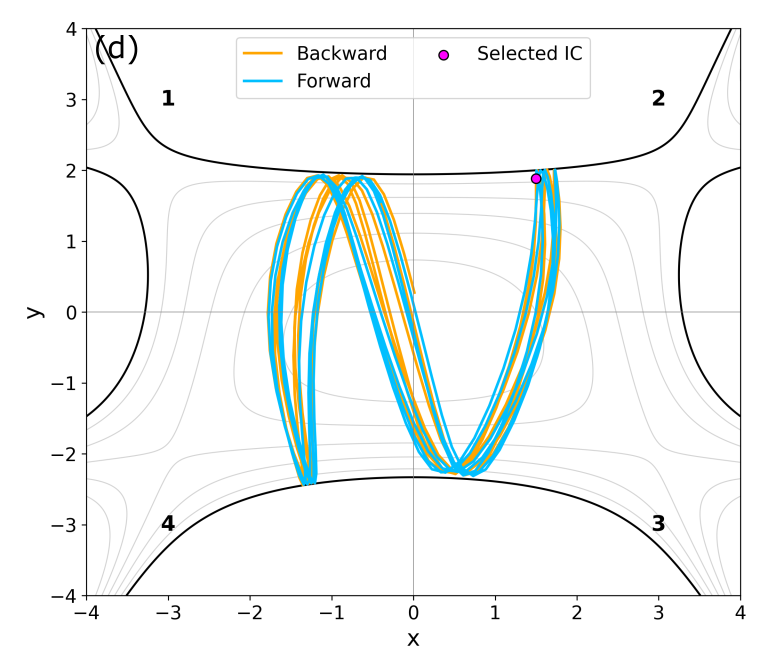}
    \caption{OFMs for the caldera potential (\eqref{eq:caldera_potential}) and orbits for the PSS defined by $y = 1.88409$, $p_y > 0$. (a) OFM for $\lambda = 1.0$, $\tau = 20$. (b) Forwards (blue) and backwards (orange) evolution of the orbit starting at the highlighted initial condition in (a). (c) OFM for $\lambda = 0.65$, $\tau = 20$. (d) Corresponding orbit from the highlighted initial condition in (c).}
    \label{fig:ofm_case1}
\end{figure*}

Fig.~\ref{fig:ofm_case1} shows OFMs for two stretching parameters, $\lambda = 1.0$ (Panels (a) and (b))and $\lambda = 0.65$ (Panels (a) and (b)), along with the corresponding forwards (blue) and backwards (orange) orbits from the selected initial condition (the magenta marker).  
For $\lambda = 1.0$ [Fig.~\ref{fig:ofm_case1}(a),(b)], the structure matches Fig.~3(a) of Hillebrand et~al.~\cite{Hillebrand2023}, with the high symmetry of the unstretched caldera producing nearly equal representation of the four channel‐to‐channel combinations.  
In this case, the saddle regions connecting the central well to each exit have similar geometries and the orbits considered to have enough energy to escape, leading to a balanced portion of initial conditions among the exits.  

In contrast, $\lambda = 0.65$ [Fig.~\ref{fig:ofm_case1}(c,d)] exhibits an expanded trapped region. This is also apparent when simulating the intermediate $\lambda$ values found in Section~\ref{sec:SupFig}.
Horizontal stretching of the potential alters the curvature and width of the saddles along the $x$‐direction, effectively lowering the barrier for orbits accessing channels~3 and~4 while narrowing the passageways to channels~1 and~2.  
As a result, the corresponding OFM shows contiguous regions drawn into the lower channels. As it expands, the inner trapped regions become more chaotic and complex, including more orbist escaping in random channels, creating colourful zones.  
This chaotic nature suggests the strong influence of the stretching parameter on the global transport network, a phenomenon also reported in other open Hamiltonian systems \cite{Wiggins1992,Ott2002}.  
The complex blending of origin–fate regions visible in both cases directly results from the stable and unstable manifolds of hyperbolic invariant sets in the phase space.  
These manifolds form the separatrices that channel orbits between exits and are responsible for the stringy patterns observed in the OFM.  
In the $\lambda = 0.65$ case, these manifolds are more twisted and closely packed, which increases the likelihood of chaotic scattering and mixed‐mode outcomes.  
Such features are closely related to the lobe dynamics framework \cite{Wiggins1992} and suggest the presence of a fractal‐like hierarchy of transport boundaries that could persist to arbitrarily fine scales \cite{Bleher1988,Viana2017}.  

\begin{figure*}[ht]
    \centering
    \includegraphics[width=0.48\textwidth]{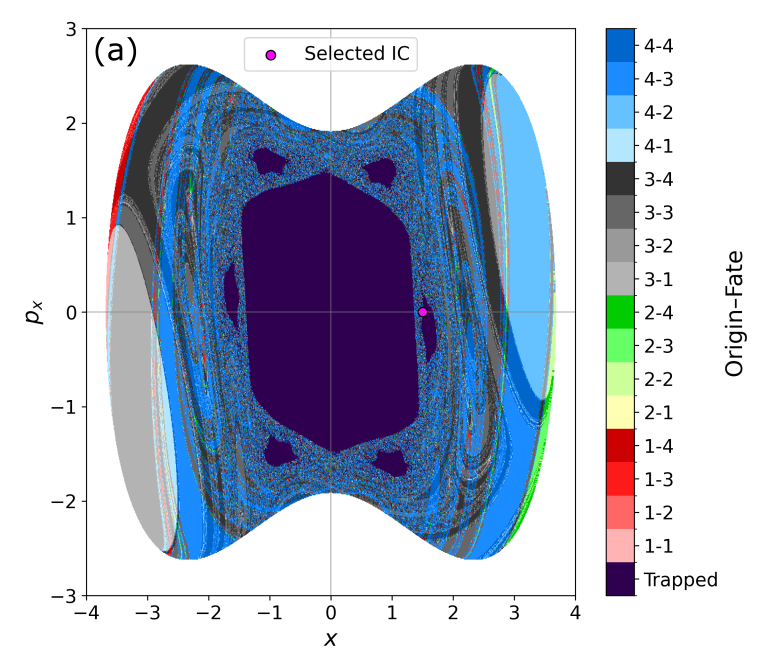}
    \includegraphics[width=0.48\textwidth]{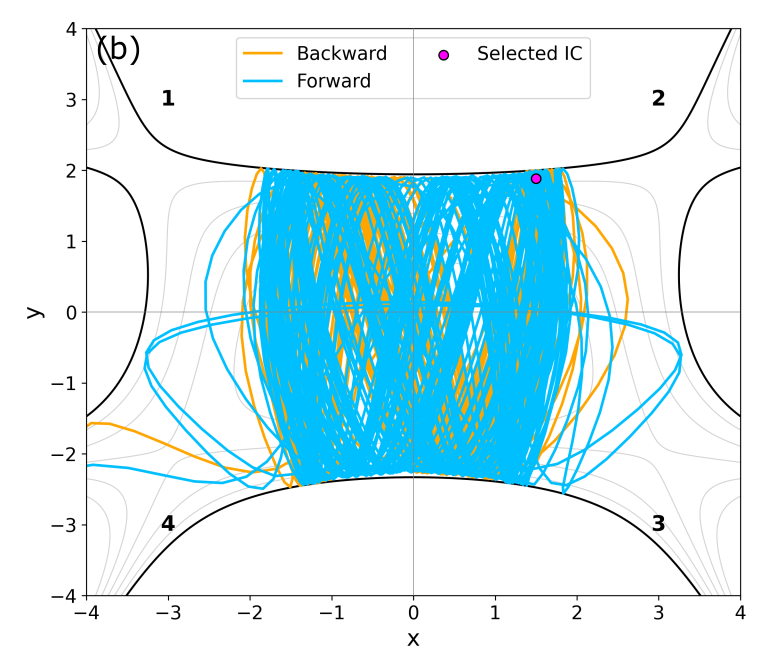}
    \caption{Long‐time integration: (a) OFM for $\lambda = 0.65$, $\tau = 500$. (b) Corresponding orbit from the highlighted initial condition in (a). Longer integration increases the fine‐scale complexity of origin–fate boundaries without altering the overall channel topology.}
    \label{fig:ofm_tau500_case1}
\end{figure*}

Although the present analysis uses $\tau = 20$, extending the integration time can further resolve the fine‐scale structure of the OFM and reduce the fraction of orbits that remain trapped within the well over the simulation window. To investigate the effect of longer integration, we also consider the $\lambda = 0.65$ case with $\tau = 500$ (Fig.~\ref{fig:ofm_tau500_case1}). 
 
Extending $\tau$ substantially increases the resolution of phase space structures, revealing more intricate origin–fate regions and sharper separatrices between transport channels.  
The increasingly fine boundaries reveal the development of thread-like structures and subtle lobe dynamics that do not appear at $\tau = 20$.

Theoretically, as the integration time $\tau$ increases, the proportion of initial conditions that remain unescaped (i.e., appear “trapped” within the potential well during the finite simulation window) should decrease. This behaviour is shown in Fig.~\ref{fig:ofm_tau500_case1}, where extending the integration time increases the structure of the origin–fate boundaries while reducing the number of trapped orbits.

Interestingly, Fig.~\ref{fig:ofm_tau500_case1} shows smaller trapped regions surrounding the larger central trapped region. These appear as circular and fractal‐like structures. This would be reminiscent of the PSS of the Hénon–Heiles system, where islands of stability are embedded within chaotic seas that themselves contain smaller islands in a loop \cite{Wiggins1992,Ott2002,Bleher1988}. Such a structure would imply that the boundaries between different origin–fate regions are not only intricate but may have an effectively infinite depth of complexity.

\subsection{Origin--Fate Maps for the PSS defined by $y = 0.3$, $p_y < 0$}

\begin{figure}[t]
    \centering
    \includegraphics[width=0.48\textwidth]{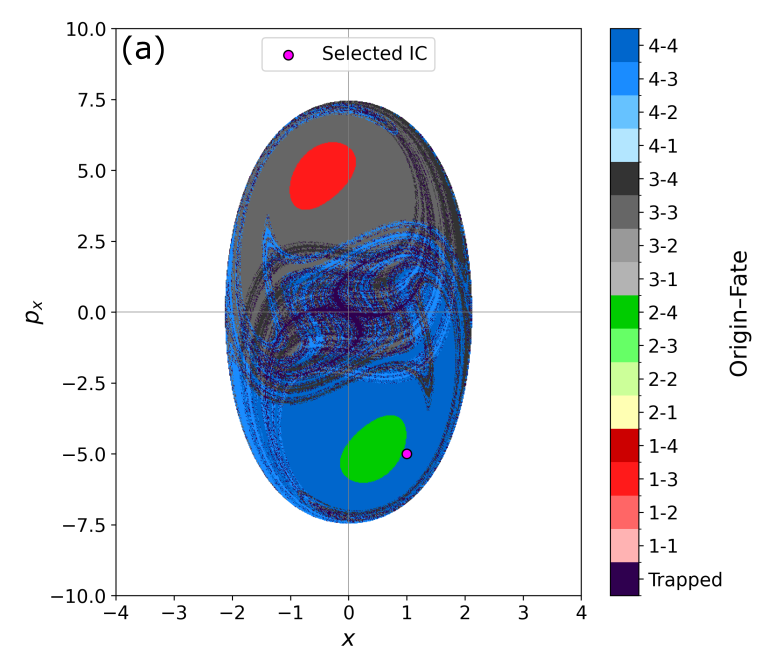}
    \hfill
    \includegraphics[width=0.48\textwidth]{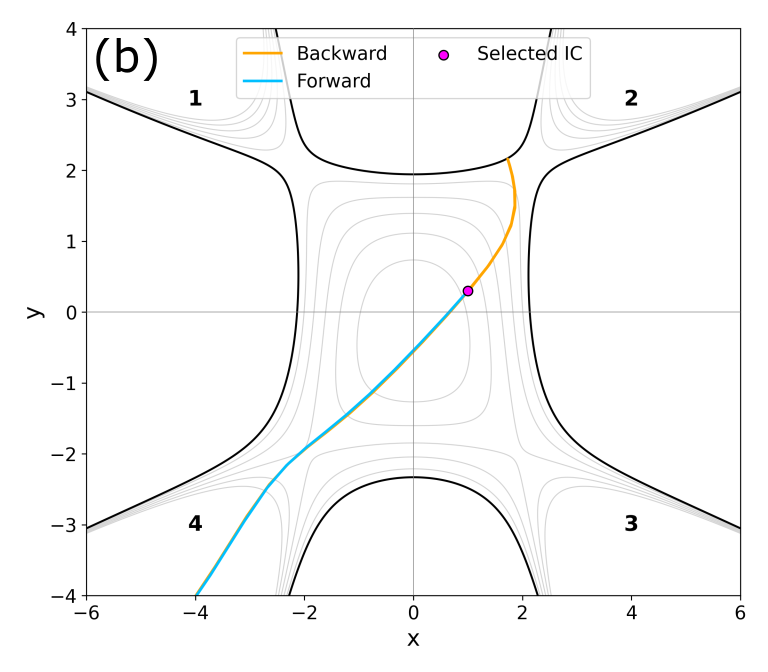}
    \vskip0.5em
    \includegraphics[width=0.48\textwidth]{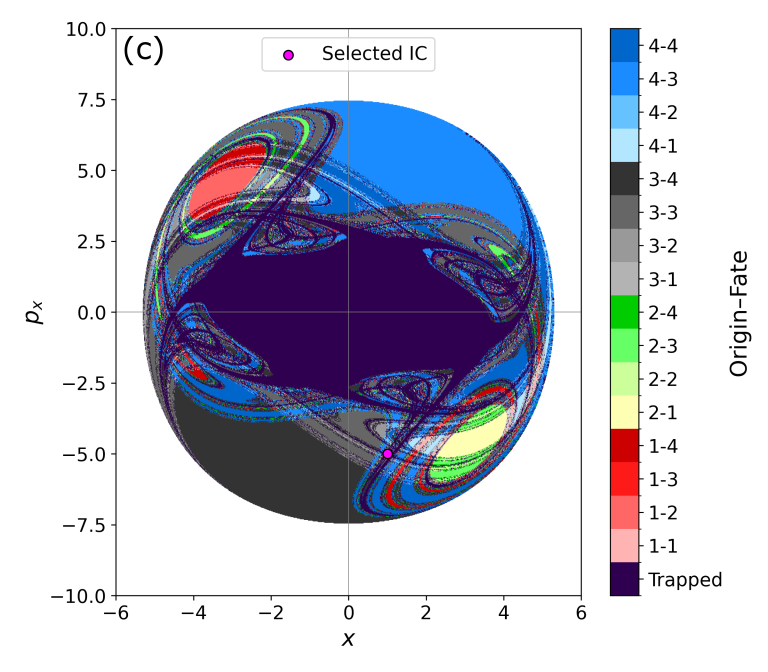}
    \hfill
    \includegraphics[width=0.48\textwidth]{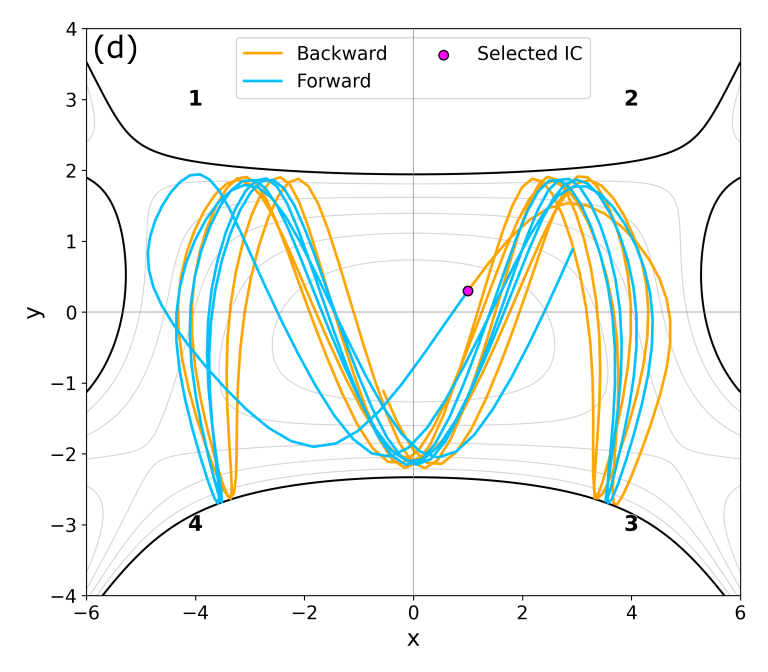}
    \caption{(a) OFM for $\lambda = 1.00$, $y = 0.3$, $p_y < 0$, $\tau = 20$. (b) Forwards (blue) and backwards (orange) orbits from the highlighted initial condition, showing perfect mirroring. (c) OFM for $\lambda = 0.40$ with the same section. (d) Corresponding orbit, showing asymmetric channel occupation, looping figure‐eight structures, and loss of forwards–backwards overlap.}
    \label{fig:ofm_case2}
\end{figure}

For an alternative OFM, we consider the PSS $y = 0.3$ with $p_y < 0$, so that orbits enter the central well from above instead of below. The computation follows Sec.~\ref{sec:methods:ofm}, where forwards and backwards integrations determine each initial condition's origin–fate classification. Unlike the case discussed in Sec.~\ref{sec:results:case1}, this choice of PSS changes the set of orbits and alters the structure of the OFM.  
In Fig.~\ref{fig:ofm_case2}, two representative stretching parameters are shown: the unstretched $\lambda = 1.00$ and the highly stretched $\lambda = 0.40$, each computed on an $800\times800$ grid of initial conditions.

In the $\lambda = 1.00$ case [Figs.~\ref{fig:ofm_case2}(a),(b)], the symmetry of the potential produces a \textit{yin–yang} like distribution of origin–fate regions, with red–green reflecting one another on opposite ends of the picture. 
The smooth and vibrant colour boundaries show vivid transitions between transport regions.  
An interesting observation is that the forwards (blue) and backwards (orange) evolution of the initial condition (magenta marker) in Fig.~\ref{fig:ofm_case2} overlap and directly trace one another. This identical mirror tracing comes from the symmetry of the Hamiltonian (\ref{eq:hamiltonian}) combined with the symmetry of the chosen initial condition, the forwards and backwards dynamics mirror each other along the same invariant manifolds but reversed.  
As a result, the time‐reversed path of the forwards integration lies exactly along the backwards orbit, which is rare.

In contrast, the $\lambda = 0.40$ case [Fig.~\ref{fig:ofm_case2}(c),(d)] shows a pronounced chaotic region and manifolds.  
Here, the horizontal stretching of the potential allows particles to escape more efficiently through different escape channels.
The origin-fate structures loop into figure‐eight‐like patterns, and the colour boundaries become thin, curves mirroring manifolds.  
In this regime, the forwards and backwards orbits from the same initial condition diverge rapidly after leaving the central well, following entirely different global routes to the exits.  
This loss of the overlap is a manifestation of chaotic transport: small differences in the manifold geometry between forwards and backwards flows lead to macroscopic separation of orbits that were initially indistinguishable in the section.

\begin{figure}[H]
    \centering    
    \includegraphics[width=0.32\textwidth]{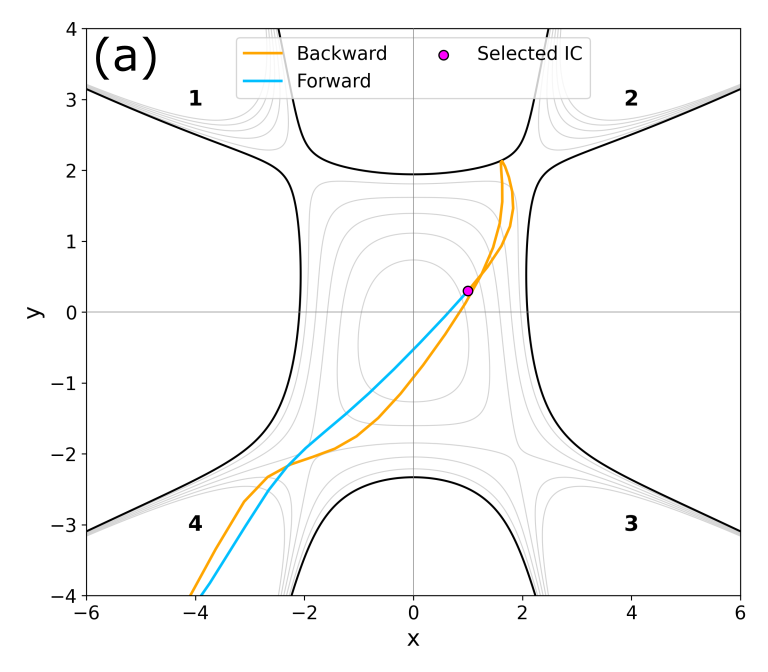}
    \includegraphics[width=0.32\textwidth]{Pics/Trajectory_0.3_-5.0_lambda_1.000.png}
    \includegraphics[width=0.32\textwidth]{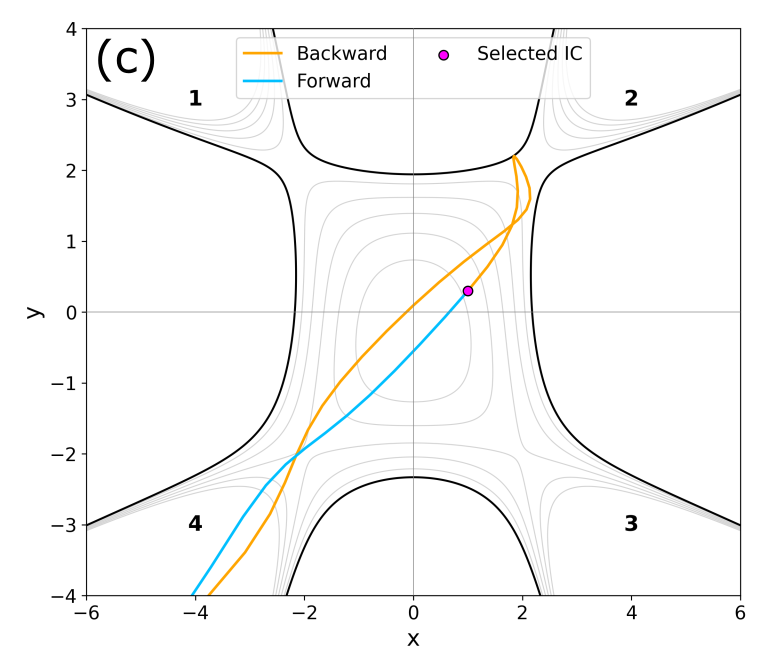}
    \caption{Forwards (blue) and backwards (orange) evolutions of the orbits starting on the PSS $y = 0.3$, $p_y < 0$, for $\tau = 20$, with $\lambda$ values of $1.02$ (a), $1.00$ (b), and  $0.98$ (c). Small changes in $\lambda$ place the initial condition in different manifold lobes, leading to different evolutions despite near‐identical starting points.}
    \label{fig:traj_overlap}
\end{figure}

To further probe the sensitivity of orbit geometry to initial momentum, Fig.~\ref{fig:traj_overlap} shows forwards and backwards evolution of orbits for three $\lambda$ values: $1.02$, $1.00$, and $0.98$.  
While all three start from almost identical phase space coordinates, their behaviour changes significantly.  
Although all three orbits originate from nearly identical phase space coordinates, their subsequent evolution changes significantly.
As $\lambda$ decreases, the geometry of the potential becomes slightly distorted, which alters the local phase space structure and shifts the orbit path.
Even a small parameter change of $\lambda = 0.02$ is sufficient to produce visibly different orbits.
This sensitivity is the hallmark of fractal well boundaries: Minute variations in initial conditions can lead to different evolutions, implying that phase space repeats at finer scales \cite{Viana2017}.
In such settings, forwards and backwards orbits fail to mirror each other and may cross in configuration space before exiting through separate channels.
This overlap–then–diverge behaviour shows that the orbits share a common unstable manifold segment before being redirected by intersecting stable manifolds.

A zoomed‐in OFM over $x \in [-4.5, 0]$, $p_x \in [2, 7.5]$ was computed using a grid of $1500\times1500$ initial conditions, focusing on the fine‐scale transport structure of the $\lambda = 0.40$ case [Fig.~\ref{fig:ofm_zoom2}].  
This region, chosen from Fig.~\ref{fig:ofm_case2}(c), reveals an extraordinary level of intricacy. Figure‐eight patterns are embedded within lobe‐like structures, which are bound by thread‐like boundaries.  
The strings correspond to manifold channels, along which transport is highly constrained and vivid.
The alternation of lobes and strings is reminiscent of the ``turnstile'' mechanism in lobe dynamics, where regions of phase space are periodically exchanged between wells from manifolds crossing one another \citep{Wiggins1992,RomKedar1990}.  
Enlarging the OFM shows smaller loops nested within larger figure‐eight shapes, suggesting the possibility of infinite fractals.  
If the resolution were increased further, one might expect to find islands of stability embedded within chaotic regions, surrounded by further layers of smaller islands.  
These features are exactly the type of boundaries that the LD gradient analysis in Sec.~\ref{sec:ld_results} is designed to uncover.

\subsubsection{Analysis of the Lagrangian Descriptors}
\label{sec:ld_results}

\begin{figure}[H]
    \centering
    \includegraphics[width=0.45\textwidth]{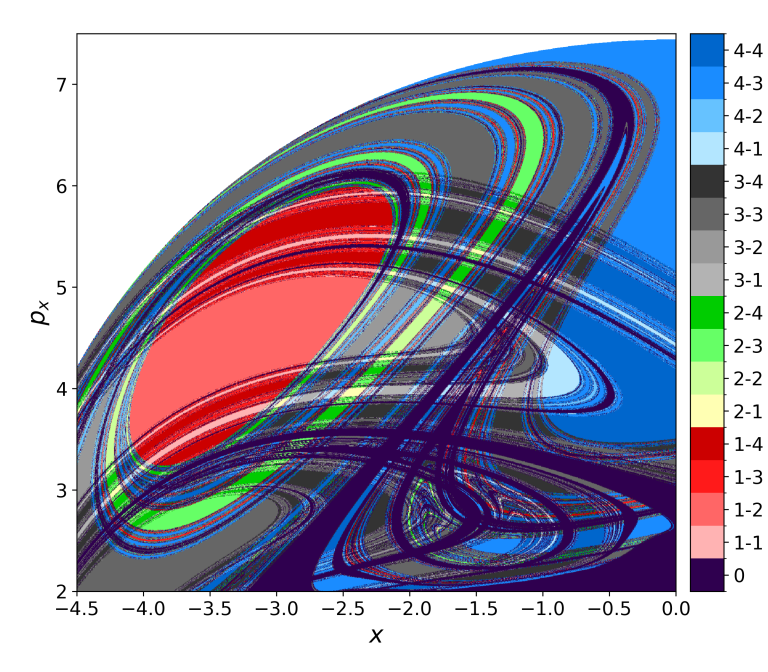}
    \caption{Zoomed view of the $\lambda = 0.40$ OFM from Fig.~\ref{fig:ofm_case2}(c) over $x \in [-4.5, 0]$, $p_x \in [2, 7.5]$, computed with $1500\times1500$ initial conditions. Fine‐scale figure‐eight patterns, lobes, and thread‐like boundaries are visible, showing complex manifold intersections.}
    \label{fig:ofm_zoom2}
\end{figure}

The zoomed OFM in Fig.~\ref{fig:ofm_zoom2} contains some of the most intricate phase space structures observed in this study, including figure-eight shapes, string-like boundaries, and lobe structures.  
To extract the invariant manifolds that govern these transportation patterns, we apply the LD methodology described in Sec.~\ref{sec:methods:ld} to the same $1500\times1500$ initial condition grid used for the OFM computation.  
The LD is computed forwards and backwards in time over $\tau = 20$, producing a scalar field $\mathcal{L}(x,p_x)$ whose sharpest ridges align with the systems stable and unstable manifolds.

\begin{figure*}[ht]
    \centering
    \includegraphics[width=\textwidth]{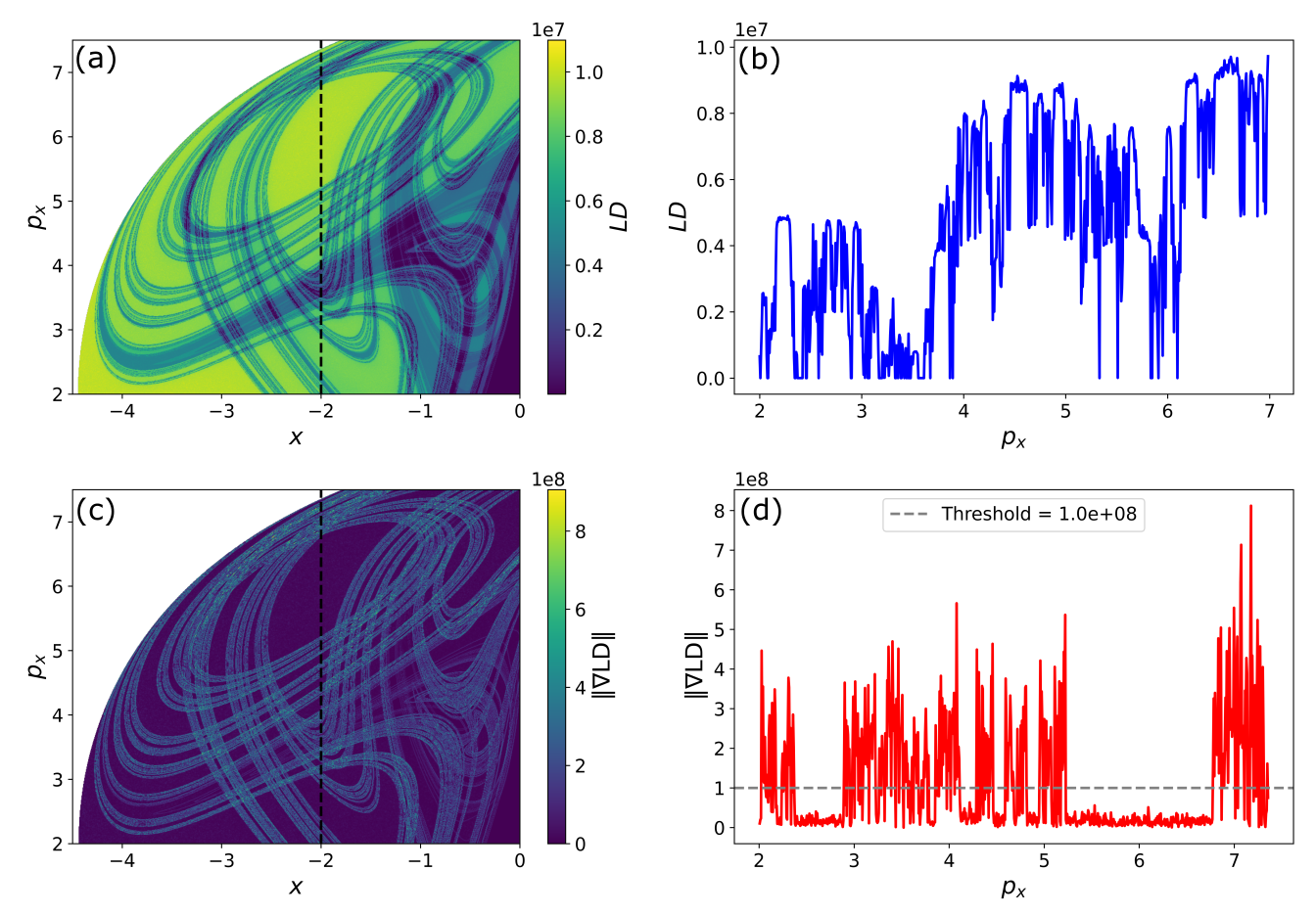}
    \caption{LD analysis on the zoomed region of Fig.~\ref{fig:ofm_zoom2}: 
    (a) LD field, (b) LD values along the $x = -2.0$ line, (c) gradient magnitude of the LD, (d) gradient magnitude along the same line. The ridges in (a),(c) correspond to invariant manifolds, while (b),(d) quantify their sharpness along the chosen cut.}
    \label{fig:ld_zoom}
\end{figure*}

Figure~\ref{fig:ld_zoom} summarises the LD analysis,  
Fig.~\ref{fig:ld_zoom}(a) shows the LD field itself: the bigger loops correspond to large transport boundaries, while finer threads are narrow manifold channels crossing the zoomed region.  
Panel~(c) displays the LD gradient, which enhances the thread features and makes the manifold skeleton more visually apparent.  
Panels~(b) and~(d) show one‐dimensional slices of the LD and its gradient magnitude, respectively, taken at the $x = -2.0$ value (marked by a dashed line in panels~a,c).  
The horizontal dashed line in panel~(d) denotes the chosen threshold (\num{1e8}), all points above that line are considered to belong to the manifold set.  
This threshold selection follows the procedure in Sec.~\ref{sec:methods:ld}, ensuring that only the sharpest and most relevant features are extracted.

\begin{figure}[H]
    \centering
    \includegraphics[width=0.49\textwidth]{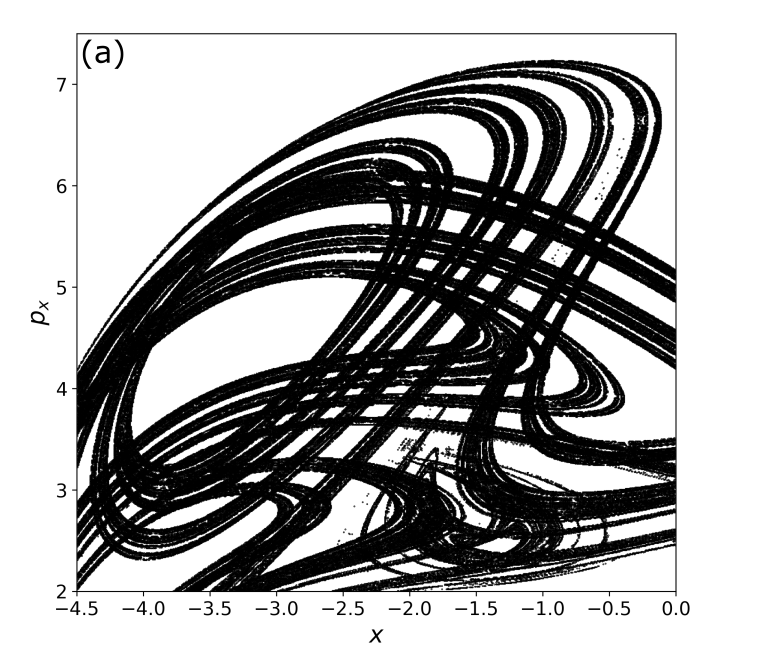}
    \hfill
    \includegraphics[width=0.49\textwidth]{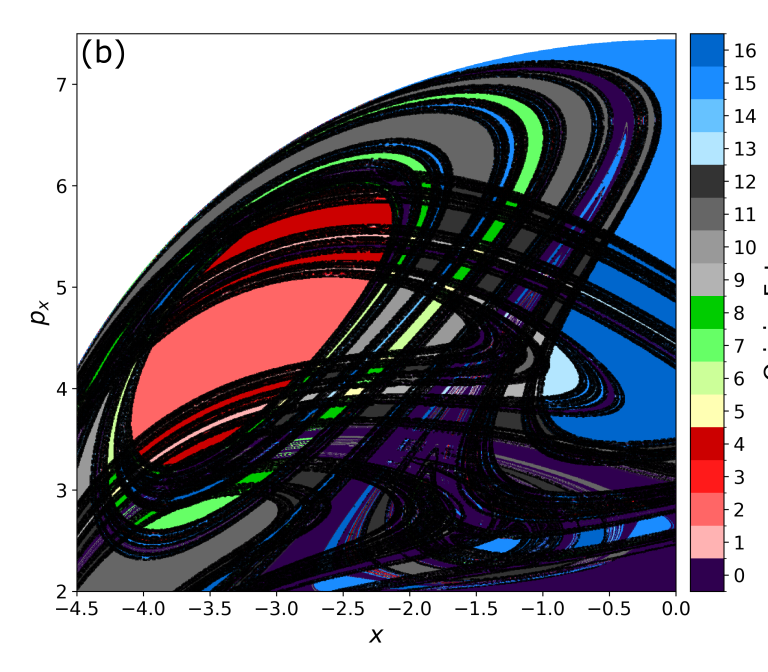}
    \caption{(a) Manifolds extracted by thresholding the LD gradient magnitude of Fig.~\ref{fig:ld_zoom}(c) i.e. points with LD gradient larger than \num{1e8} are shown. 
    (b) Same manifolds overlaid on the zoomed OFM of Fig.~\ref{fig:ofm_zoom2}, showing near‐perfect correspondence between manifold curves and origin--fate boundaries.}
    \label{fig:ld_manifolds}
\end{figure}

Applying this threshold to the entire gradient magnitude map yields the manifold set shown in Fig.~\ref{fig:ld_manifolds}(a).  
These manifolds form a complex web of intersecting curves, outlining the boundaries between dynamically distinct regions of phase space.  
When these extracted manifolds are overlaid on the zoomed OFM in Fig.~\ref{fig:ld_manifolds}(b), a remarkable alignment is observed: the colour boundaries in the OFM, separating different origin–fate classifications—coincide almost exactly with the extracted manifold curves.  
This confirms that the intricate lobe and string‐like structures seen in the OFM are generated and sustained by the underlying invariant manifold network.

Compared to the LD manifold plots in \cite{Hillebrand2023}, which focused on a different region of phase space, our zoomed‐in analysis reveals a higher density of manifold intersections and a greater variety of enclosed lobe shapes.  
This is a direct consequence of both the smaller domain size and the higher spatial resolution used here.  
The alternation of figure of eight loops and thread‐like connectors is consistent with the ``turnstile'' picture of lobe dynamics. 
The presence of manifolds that weave through multiple colour regions in the OFM also explains the strong sensitivity to initial conditions observed in the orbit overlap in Fig.~\ref{fig:traj_overlap} as tiny changes can move an initial condition across a manifold boundary, completely changing its channel.

The strong correlation between LD manifolds and OFM boundaries highlights the similar nature of the two approaches:  
the OFM provides a visualisation of transportation outcomes for a given system, while the LD gradient can be used to isolate the invariant structures that enforce these boundaries.  
Together, they give a clear consistent image of the transport mechanisms operating in the caldera potential.

\section{Conclusion}
\label{sec:conclusions}

In this study the OFM methodology was applied, in combination with LDs, to investigate phase space transport in a two-dimensional caldera potential. The four-exit caldera system with a tunable stretching parameter $\lambda$ serves as an ideal model for studying the correlations between chaotic scattering, islands of stability, and invariant manifold structures in Hamiltonian dynamics.

Our OFM results for the surface $y = 1.88409$, $p_y > 0$ reproduced, for $\lambda = 1.0$, the highly symmetric pattern reported in \cite{Hillebrand2023}, with balanced channel accessibility and evenly distributed mixed regions. In that case, the forwards and backwards integrations produced an image in which nearly all orbits escaped the potential well, reflecting the overall openness of the unstretched potential. As $\lambda$ decreased, the chaotic behaviour increased and channel accessibility became highly uneven as trapped zones enlarged, particularly at $\lambda = 0.65$, where horizontal stretching favoured transport into the lower channels. Extending the integration time from $\tau = 20$ to $\tau = 500$ revealed more intricate structures and reduced the set of trapped orbits.

The second OFM configuration on the PSS defined by $y = 0.3$, $p_y < 0$ showed markedly different structures, reflecting reverse entry into the central well. For $\lambda = 1.0$, the OFM displayed a yin–yang–like pattern with almost all orbits escaping, and the forwards and backwards orbit paths overlapped nearly perfectly (a rare time-mirrored behaviour). For $\lambda = 0.4$, the OFM exhibited rich figure-eight loops and alternating colour strands associated with intertwined manifolds; in this regime the forwards and backwards orbits no longer overlapped. Integrating three nearby initial conditions highlighted the system’s extreme sensitivity: tiny changes in starting points led to entirely different orbit paths and exit channels.

The zoomed OFM analysis, at high resolution ($1500\times1500$ grid), revealed a dense web of string-like structures, lobe patterns, and transport barriers. To interpret these features, the LD fields and their gradient magnitudes on the same domain were computed. By keeping points whose LD gradient exceeded a fixed threshold, the manifold skeleton was extracted which, when overlaid on the OFM, showed near-perfect correspondence with origin–fate boundaries. This confirms that the colour boundaries in the OFM are generated by the invariant manifolds that act as barriers to transport in the caldera. Compared to the manifold extractions in \cite{Hillebrand2023}, our focus on a more intricate region exposed a higher density of manifold intersections and finer-scale transport channels.

Together, the results demonstrate how the OFM and LD approaches provide complementary, in-depth perspectives on transport in Hamiltonian systems. The OFM gives an outcome-based classification of phase-space initial conditions, while the LD computation exposes the invariant manifold structure that organises those outcomes. By applying both methods across multiple surfaces of section, stretching parameters, and resolutions, we build a coherent picture of transport in the stretched caldera potential—from predictable scattering at $\lambda = 1.0$ to more chaotic, sensitive transport at $\lambda = 0.4$.

Future work could extend this framework to even longer integration times to quantify the decay of trapped sets, and to test whether OFM boundaries exhibit fractal structure at progressively smaller scales. In addition, applying the same methodology to other multi-channel potentials, or to time-dependent Hamiltonian systems, could reveal new transport mechanisms beyond those identified here.

\section{Supplementary Figures and Animations}
\label{sec:SupFig}
\centering Supplementary videos and images are available online via \cite{moser2025ofmdata}.
\section{Acknowledgements}
\label{sec:acknowladge}
The author would like to thank Prof.~Haris Skokos for his supervision, guidance, and invaluable feedback throughout the course of this project. 
Gratitude is also extended to the Nonlinear Dynamics and Chaos Group at the University of Cape Town for providing a stimulating research environment and insightful discussions. 
Special thanks to the developers of the ABA864 symplectic integrator and the open-source scientific computing community, whose tools made the numerical experiments possible. 
Finally, the author acknowledges the inspiration provided by the work of Hillebrand et~al.~\cite{Hillebrand2023}, which served as the foundation for the present study.

\bibliographystyle{apsrev4-2}
\bibliography{bib} 

\end{document}